\documentclass{article}

\usepackage[english]{babel}
\usepackage[margin=3cm]{geometry}
\usepackage{newlfont}
\usepackage[T1]{fontenc}
\usepackage{tgpagella}

\usepackage{authblk}
\usepackage{float}
\usepackage{csquotes}
\usepackage[
noibid,
backend=biber,
authordate,
natbib,
]{biblatex-chicago}
\usepackage{graphicx}
\usepackage[font={small,it}]{caption}
\usepackage{hyperref}
\hypersetup{colorlinks=true, allcolors=blue}
\usepackage[table,xcdraw]{xcolor}
\usepackage{fancyhdr}
\usepackage{comment}
\usepackage{longtable}

\title{A First Ontological Model for the Description of the Art Market in the Semantic Web}
\author{Manuele Veggi\thanks{National PhD in “Heritage Science": University La Sapienza of Rome / CNR - Institute of Heritage Science; email \href{mailto:manuele.veggi@uniroma1.it}{manuele.veggi@uniroma1.it}.}}
\date{}

\begin{document}

\maketitle

\section*{Abstract}
This dissertation presents the first version of a project at the Fondazione Federico Zeri, aimed at modelling the art market starting from the recognition of the peculiarities of this sector and relying on the data collected by this institute during its research activities on its documentary collection. Specifically, this study describes the development of an ontology, able to describe agents, events and sources which define the art market and enable its investigation. The recognition of existing conceptual models is hence followed by the description of the adopted methodology, based on the protocol SAMOD. The central section provides a general overview of the final ontology, integrating the results of a preliminary study. Lastly, the appendix lists motivating scenarios, examples and competency questions collected during the first SAMOD iterations, as well as a first alignment with existing models.

\section*{Keyword}
Ontology engineering $\cdot$ Digital Art History $\cdot$ Art Market $\cdot$ Knowledge Representation $\cdot$ Semantic Web

\section*{Premise}

This research, presented as end-of-study dissertation at Collegio Superiore of “Alma Mater Studiorum" University of Bologna (a.y. 2022/2023, supervisor: Prof. Iannucci, co-supervisor: Prof. Peroni), is aimed at a first semantic modelling of the art market. The first part of this project was carried out as a curricular internship at the Fondazione Zeri - University of Bologna (Jan. - Mar. 2023, tutor: Prof. Mambelli): the collaboration with the staff of this research centre allowed me to collect the necessary requirements to model this domain. The outcome of the first exchange with the collaborators of the Foundation were collected in a preliminary study, which was written together with Prof. Mambelli and was eventually published on the journal \textit{Umanistica Digitale} \parencite{veggiModellingArtMarket2023}.

The following phases of the research were hence centred on the development of the ontology and its documentation following this initial overview: as a result, several sections of the current paper are hence drawn from this article, with small or no modifications. More relevant changes have been instead implemented in the central paragraph (see \ref{sect_zamo_mod}), which describes the actual configuration of the ontology and how it differed from the preliminary analysis. Therefore, an essential integration of this paper are:
\begin{itemize}
    \item the material hosted in ZAMO GitHub repository: \url{https://github.com/fondazerimv/zamo};
    \item the final documentation, available at \url{https://w3id.org/zeri/ontology/zamo}.
\end{itemize}

\newpage
\section{Introduction}
Among the many fields of research related to art history, art market studies is one that has seen a major increase in recent years. After focusing on the study of artists' personalities, contributing to the reconstruction of their catalogues, and defining a shared art historical canon, scholars have recently returned the history of works as material objects to the center of their interest. The movements of paintings, sculptures, and other artistic items; their confluence in different collections that reflect specific tastes; the dispersal and recomposition of cultural heritage in areas different from those in which they were produced; the commercial strategies of the agents who facilitated this network of trade: all these have become increasingly investigated topics. 

Many initiatives and interdisciplinary projects about the art market and history of collecting were promoted by research centers and a rich literature was produced, including specialized journals and entire book series. Among them, at least the Zentralarchiv für deutsche und internationale Kunstmarktforschung (ZADIK); the Forum Kunst und Markt, with its Journal for Art Market Studies; the Project for the Study of collecting and Provenance of the Getty Research institute; the Antique dealers project of the University of Leeds, are worth mentioning (for an overview, see \cite{caraffaIntroductionPhotographyArt2020}). Moreover, the Bloomsbury Art Markets database, including numerous entries about protagonists of the international art market since 1900, has been recently launched with the aim of becoming a reference point for scholars in the field (\cite{bloomsburyBloomsburyArtMarketsn.d.}).

The Federico Zeri Foundation has actively participated in this debate, with a focus on the art market in Italy in the 19th and 20th centuries. In addition to promoting conferences and publications (\cite{bacchi2020capitale}), it has recently supported two research grants aimed at collecting unpublished information on dealers and collectors linked to the archive and the personal history of its founder.

As it is well known, the Fondazione preserves the documentary collections of Federico Zeri (1921- 1998), who was one of the leading art historians of the last century and an infallible connoisseur. They comprise a photo archive composed by 290,000 photographs of artworks and monuments, a library made of 46,000 books, and a collection of 37,000 auction catalogues, the largest of its kind in Italy. The scholar has used these resources as indispensable working tools to carry out his work as an independent scholar. In his career, he has been an adviser to dozens of art galleries, antiquaries, collectors, and auction houses, guiding them in their respective purchases and sales on the market. This activity is reflected in the handwritten annotations on the photographs and volumes. Among them, more than 500 names of individuals active in the market during the last two centuries were retrieved. How to systematize, normalize, and deepen data about these subjects was one of the main aspects of the cataloging project of the documentary collections (\cite{mambelliRisorsaOnlineStoria2014}).

Despite the studies and resources cited above, the Foundation realized how difficult it was to retrieve complete and detailed information about these figures using traditional art historical research methodologies. The nature of the art market entails a number of resistances to the dissemination of data on buyers, dealers, and sale prices of works. The lack of reliable data about transactions has proven to be the first in a series of aspects that characterize this domain and distinguish it from other business sectors.

First, those of antique dealers are often family-run businesses in which grandparents, children and grandchildren frequently use the same names: this results in extreme difficulty in figuring out to which different individual certain information refers. On the other hand, the corporate name of a given gallery may change several times over time, for example, as a result of the opening of new branches or the establishment of more or less stable joint ventures with other partners. Alongside the figure of the dealer revolves a wide network of advisers and collaborators. These include restorers, informants, and scholars with expertise in certain areas of art historical production, or in specific periods and styles, whom antiquarians or auction houses use to determine the quality and value of particular pieces. They may assist in writing files and catalogues, organizing sales, promoting dealers to collectors or museum directors. Being able to reconstruct and represent this network can contribute significantly to understanding the contexts and mechanisms of the art market.

Another typical aspect concerns what appears to be the very core of the domain: the transaction. The price of the works offered for sale depends not on their intrinsic characteristics but on a number of external elements that concur to determine their commercial value: first of all, on the attributions made by more or less authoritative art historians that accompany the pieces; then on their collecting history and state of preservation; again, on the fortune that a given style or artist comes to have at a given historical moment, in parallel with the evolution of fashion and taste, and so on.

Given the difficulty of accessing the antiquarians' personal archives, the study of the art market involves the use of other, sometimes non-canonical sources: trade guides, narrative and autobiographical texts, commercial magazines of the time, but, above all, the photographs of the works put up for sale initially addressed to advisors and potential clients, which have since been deposited in the photographic collections of scholars or art-historical institutions (Caraffa and Bärnighausen 2020). It is no coincidence that it was from the cataloguing projects of these archives that the need emerged to investigate and systematize information on the subject. The last type of source to be considered are the direct testimonies of market agents and their successors (children, grandchildren, colleagues etc.). The research promoted by the Fondazione involves extensive use of direct interviews and oral testimonies given by these individuals, which researchers are collecting and transcribing.

The ontology ZAMO (Zeri Art Market Ontology), presented here in its first version, aims to represent these and other specific aspects of the art market domain, and aspires to become a reference model for anyone interested in structuring, enhancing, and integrating scholarly data related to this field.

\section{State of the Art}
The construction of an ontology for the art market domain required at first a recognition of this multifaceted research field, in order to identify a selection of relevant projects concerning the description of objects, agents, and transactions typical of this domain.

As the project mainly deals with cultural heritage data, the principal reference ontology is CIDOC- CRM, the conceptual model proposed by ICOM which is a standard for the semantic description of pieces of information in this domain. One of the main peculiarities of this project is the relevance of the notion of event. Indeed, this approach permits “a more accurate view of the past or current life history of a cultural object, [...] interprets more effectively history and especially heterogeneous and complex information resources that [...] need to be linked and interpreted in order to capture knowledge” \parencite[1]{doerrDocumentingEventsMetadata2006}.

In addition, this class also permits to model one of the core elements of the art market, which is the economic transaction. The solution proposed by CIDOC-CRM puts forward a streamlined definition of this phenomenon, which is mainly corresponding to a change either in the possession or in the custody of a \verb|crm:E18_Physical_Thing| instance. Since this model already defines all the main pieces of information usually used by researchers in the art historical domain, further economic factors described by other ontologies and conceptual models \parencite[e.g. see][]{guizzardiCoreOntologyEconomic2020} arethus not considered.

In this perspective, also the final ontology is centered on this class, which enables us to link different modules conceived for the description objects, documents, agents. While the first two can still be defined by CIDOC-CRM (specifically through the class \verb|crm:E18|), the internal dynamics within art galleries are difficult to describe with the ICOM standard model. Consequently, an integration with another module, focused on the structure of organizations, has been deemed necessary.

Reynolds already proposed The Organization Ontology (Org, prefix: \verb|org:|), which is an effective tool to describe the roles and position within a company, the changes in its major characteristics, as well as modification in the personnel (\cite{reynoldsOrganizationOntology2014}). Moreover, this model shares relevant features with CIDOC-CRM class hierarchy. In particular, the class  \verb|org:ChangeEvent|, through which modifications of the studied company can be asserted, can be interpreted as a better specification of \verb|crm:E5_Event|. Besides, Org integrates a widely used pattern in modelling the agent, which is actually an imported class from FOAF (\url{http://xmlns.com/foaf/0.1/}, prefix: \verb|foaf:|) and which can be aligned with \verb|crm:E39_Actor|. Org main class, org:Organization, is defined as a subclass of foaf:Agent, on the same hierarchical level as the other two subclasses \verb|foaf:Person| and \verb|foaf:Group|.

The enhancement of CIDOC-CRM model for the description of sources and evidences has been instead already suggested by the project \textit{Zeri\&LODE}, which serialized in linked open data the metadata of the photographs of the Zeri Photo Archive related to the artworks of the Italian Cinquecento. Indeed, this initiative translated into two complete ontologies the \textit{Scheda F} (F Entry, for photographs) and \textit{Scheda OA} (OA Entry, for artworks) defined by Italian ICCD (Central Institute for Cataloging and Documentation of the Ministry of Culture), thanks to which it is also possible to express main bibliography connected to the catalogued item.  CiTO (\cite{peroniFaBiOCiTOOntologies2012}, prefix: \verb|cito:|) was hence reused to quote sources to reconstruct the attribution of an artwork, while HiCO (\cite{daquinoHistoricalContextOntology2015}, prefix: \verb|hico:|) was imported “to describe the interpretation process relative to subjective attributions” \parencite[4]{daquinoEnhancingSemanticExpressivity2017}. Lastly, an additional solution for the modelling of historical interpretations and archives  (see \textit{infra}) is provided by Italian national reference ontology of this domain, ARCO (\url{http://wit.istc.cnr.it/arco}).

\begin{figure}[ht!]
  \includegraphics[width=1\textwidth]{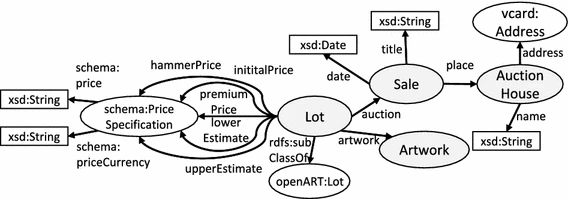}
  \caption{Excerpt of the art auction ontology, sales data \parencite[from][]{filipiakQuantitativeAnalysisArt2016}}
  \label{fig:fpk2}
\end{figure}

As the final ontology can be aligned with some of these existing modules, this brief overview serves as premise for the analysis of more specific projects, assessing both the degree of interoperability with community standards and the ability of modelling relevant information for art historical research. For instance, an interesting endeavor in modelling the art market has been done by Filipiak and colleagues, which emphasizes mainly the economic characteristics of an auction in the art market \parencite[][]{filipiakQuantitativeAnalysisArt2016}. The quoted conference paper displays two main excerpts of the ontology, reported in Fig. \ref{fig:fpk1} and Fig. \ref{fig:fpk2}:

 \begin{figure}[ht!]
  \includegraphics[width=1\textwidth]{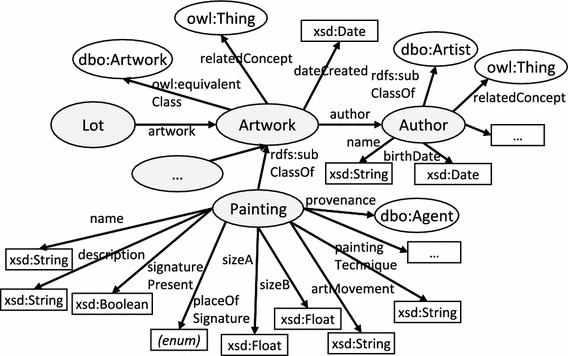}
  \caption{Excerpt of the art auction ontology, sales data \parencite[from][]{filipiakQuantitativeAnalysisArt2016}}
  \label{fig:fpk1}
\end{figure}

The latter fragment is particularly relevant as it stresses the importance of not directly linking the artwork to the sales, rather the model relies on the intermediate class of lot, the “central entity” of the shown part of the ontology, in which “a particular item is offered for sale” \parencite[6]{filipiakQuantitativeAnalysisArt2016}. Nonetheless, the project seems to consider only a specific situation, consisting of artworks by a known author (see Fig. \ref{fig:fpk2}), omitting other crucial practices of this domain (such as evaluations or expertises), which are instead important in the definition of the exchange value of the single object.

The current state of the art hence now lacks a consistent and comprehensive data model which enables the interaction between current standard modules used to describe objects, event, and agents. On top of that, the new ontology will be indeed asked to define both relevant practices of the art historical domain and to state relevant archival and bibliographical sources, in order to support and streamline future researches of this sector.

\section{ZAMO, Zeri Art Market Ontology}

\subsection{Methodology}
\label{sect_methodology}

As said, the “Zeri Art Market Ontology” is intended to build up meaningful relationships among different elements of the art market and the art-historical scholarships. Such a broad ontology hence needs to be subdivided into minor modules; in particular, three main domains have been defined.

The first one is focused on the agent (i.e., the art dealer) and describes the possible relations connecting a single person to an organization (e.g., the art gallery or the auction house). The second module is focused instead on the artwork and the agent in the context of the transaction: the various kinds of economic exchanges are taken into consideration and special attention has been paid to the analysis of value attribution processes (expertise, such as authorship attribution, and value proposition). Lastly, the third module scrutinizes the documentary sources on which art historians rely to reconstruct the activity of art dealers. As discussed in the following paragraphs, they consist of either single objects or entire curated holdings, such as archives.

The development of this ontology has been carried out following the SAMOD methodology \parencite[][]{peroniSAMODAgileMethodology2016}{}{}. Indeed, this acronym stands for Simplified Agile Methodology for Ontology Development and is based on an iterative process in which a “domain expert” (DE) and “ontology engineer” (OE) collaborate to define motivating scenarios, which are small problems defined in natural language and associated to a set of informal and likely examples. After this first step, competency questions are formulated. On this basis, the OE develops a standalone modelet and refractors it to check “formal” and “rhetorical” requirements, that is mainly to verify the absence of inconsistencies and to compare the solutions of the SPARQL competency questions with the expected results.\footnote{See SAMOD guidelines \parencite{peroniSAMODAgileMethodology2016} for a more detailed description of this methodology.}

 \begin{figure}[ht!]
  \includegraphics[width=1\textwidth]{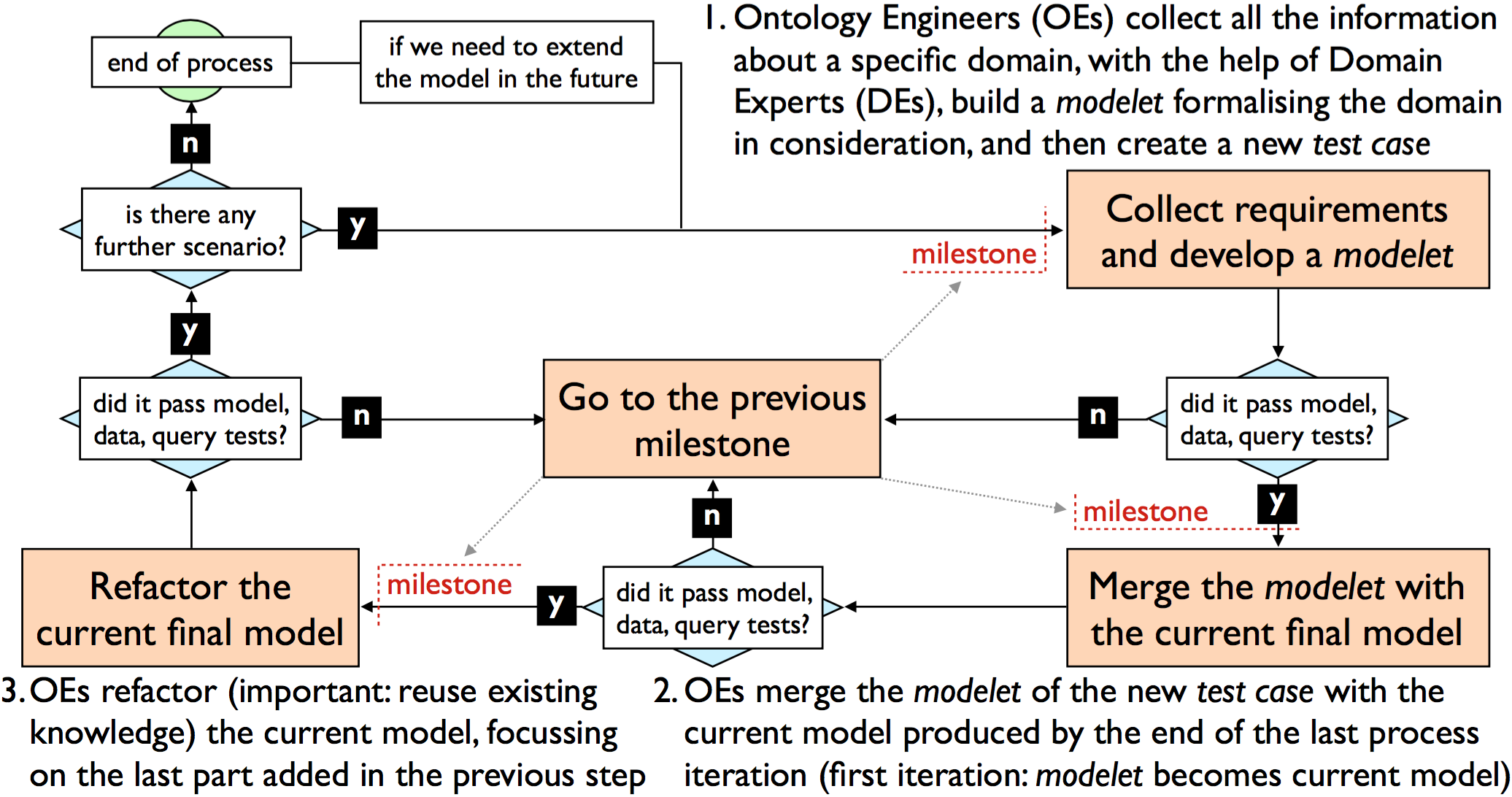}
  \caption{ SAMOD Methodology after \parencite{peroniSAMODAgileMethodology2016}}
  \label{fig:samod}
\end{figure}

These reflections serves as guidelines in the proper construction of the ontology, whose responsibility lies entirely on the OE. More specifically, for the development of ZAMO, \textit{modelets} have been firstly drawn with the software \textit{yEd}\footnote{yWorks. textit{yEd}. \url{https://www.yworks.com/products/yed}.} using the graphical framework \textit{Graffoo}.\footnote{Peroni, Silvio. \textit{Graffoo}. \url{https://essepuntato.it/graffoo/} (see \cite{falcoModellingOWLOntologies2014}).} The model has been later converted into .owl file with the tool \textit{Ditto}\footnote{Peroni, Silvio. \textit{DITTO}. \url{https://essepuntato.it/ditto/} \parencite[see][]{gangemiDiTTODiagramsTransformation2013}.} and populated in \textit{Protégé}.\footnote{Stanford University, School of Medicine, Stanford Center for Biomedical Informatics Research. Protégé. \url{https://protege.stanford.edu/}.} Fulfilment of formal requirements have been checked both in this software, with the reasoner \textit{HermiT},\footnote{Oxford University, Department of Computer Science, Data and Knowledge Group - Knowledge Representation and Reasoning. \textit{HermiT OWL Reasoner}. \url{http://www.hermit-reasoner.com/}. Reference version: HermiT 1.4.3.456.} and with \textit{OntOlogy Pittfall Scanner (OOPS!)}.\footnote{Ontology Engineering Group. \textit{OntOlogy Pitfall Scanner}. \url{https://oops.linkeddata.es/} \parencite[see][]{poveda2014oops}.} After populating the ontology with the example dataset, the plug-in \textit{SPARQL Query}\footnote{Redmond, Timothy. \textit{SPARQL Query}. \url{https://protegewiki.stanford.edu/wiki/SPARQL_Query}.} installed in \textit{Protégé} has been used to verify the correctness of the answers to SPARQL-format competency questions. On the contrary, in test case phase, rhetorical requirements of model, data, and query tests (\textit{bag of test cases}) have been checked manually. Table \ref{table:1} sums up the used design tools, associating them with the corresponding step of the SAMOD methodology. As foreseen by SAMOD methodology, modelet merging and milestones releasing have been carried out only if the requirements stated by the bag of test cases have been met.

\begin{table}[H]
\resizebox{\textwidth}{!}{
\begin{tabular}{|l|l|}
\hline
\rowcolor[HTML]{C0C0C0} 
\textbf{SAMOD Step} & \textbf{Tool} \\ \hline
\begin{tabular}[c]{@{}l@{}}Building of the modelet\\ Graphical representation\\ OWL serialization\end{tabular} & \begin{tabular}[c]{@{}l@{}}Graffoo\\ DITTO\end{tabular} \\ \hline
Model test, formal requirements & \begin{tabular}[c]{@{}l@{}}Prot\'{e}ge\'{e} – HermiT Reasoner\\ OOPS!\end{tabular} \\ \hline
Building of the dataset according to the developed modelet & Prot\'{e}ge\'{e} \\ \hline
Data test, formal requirements & Prot\'{e}ge\'{e} – HermiT Reasoner \\ \hline
Query test, formal requirements & Prot\'{e}ge\'{e} – SPARQL Query \\ \hline
\end{tabular}}
\caption{Use of existing softwares and APIs in the development of the final ontology according to SAMOD workflow}
\label{table:1} 
\end{table}

After the completion of all SAMOD iterations, a first alignment have been proposed. This task has been carried out through punning. This alignment strategy has been admitted in the transition from OWL 1 to OWL2 (see \cite{owlworkinggroupPunning2007} and \cite{grauOWLNextStep2008}) and enables ontology engineers to meta-model classes, i.e. that “an URI can denote a class and an individual at the same time" \parencite[][]{lamparter2007preference}{}{}. In this way, the alignment knowledge graph can reuse the Simple Knowledge Organization System (SKOS, \url{https://www.w3.org/2009/08/skos-reference/skos.html}, prefix \verb|skos:|): classes and properties can be instantiated as individual of type \verb|skos:Concept| and the proper alignment can be asserted through SKOS Mapping properties (see the documentation at \url{https://www.w3.org/TR/skos-reference/#mapping}). This solution have been chosen as it enables OE to assert additional relationships between entities: as an example, through the \verb|skos:relatedMatch| it is possible to express a generic associative mapping relationship between two concepts, without asserting a hierarchical link (e.g. subclass or super-class) or the strict correspondence of \verb|owl:equivalentProperty| or \verb|owl:equivalentClass| (see the use of these mapping properties in final appendix).

Lastly, the creation of the documentation in .HTML format has been carried out through the service LODE (\url{https://essepuntato.it/lode/}) \parencite[][]{peroni2012making}{}{} (with minor manual corrections on the source HTML output page).

\subsection{Modules}
\label{sect_zamo_mod}

\begin{figure}[H]
\centering
  \includegraphics[width=1\textheight, angle=270]{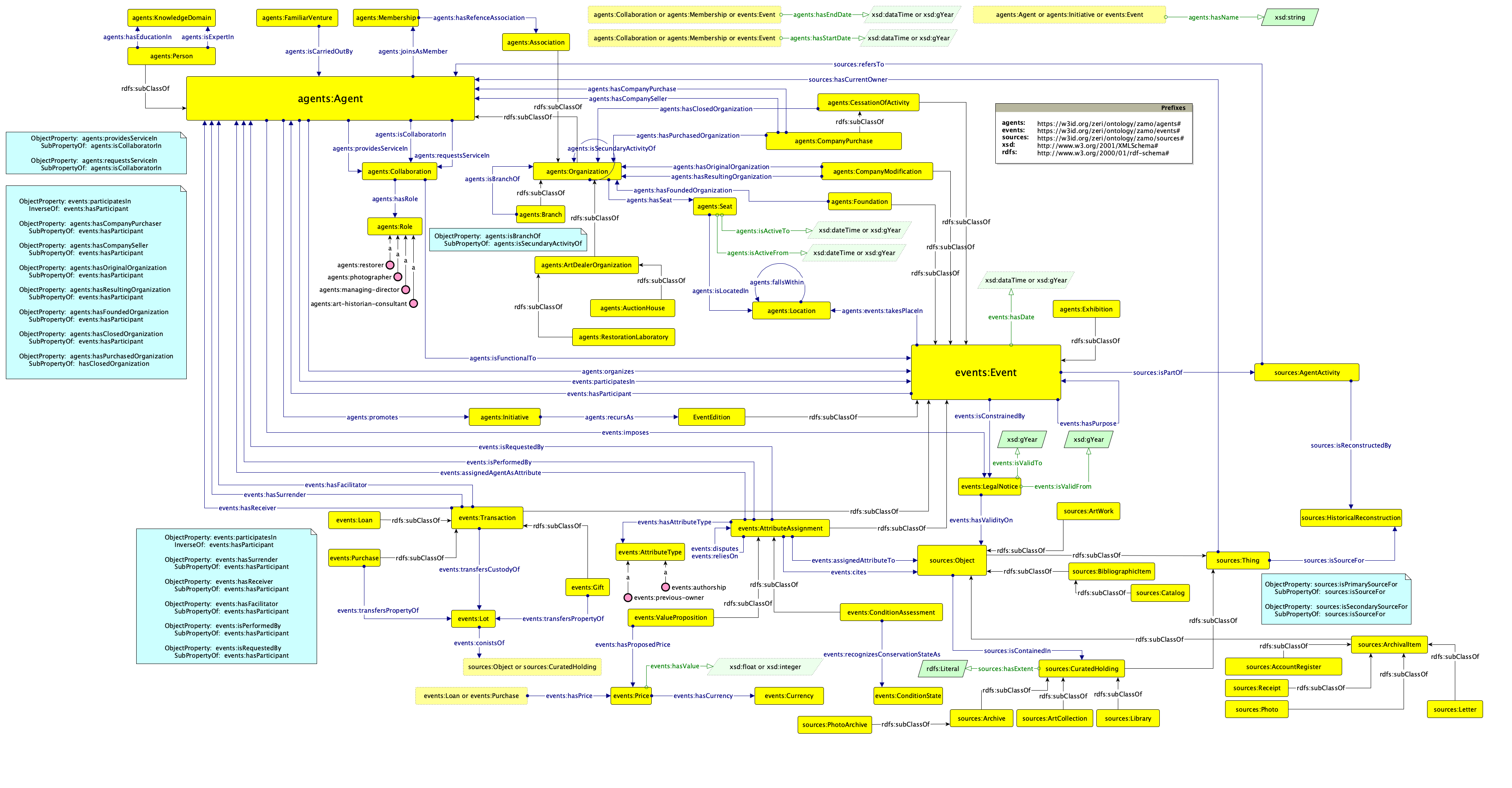}
  \caption{Conceptual Model containing all the classes, properties, and individuals in ZAMO}
  \label{fig:fullzamo}
\end{figure}

As anticipated in the previous section, Zeri Art Market Ontology is articulated into three different modules describing the most relevant agents, events and sources for the historical reconstruction of this domain. A general overview of all the asserted classes, object and data properites, and individuals can be seen in Fig. \ref{fig:fullzamo}.

As seen, the subdivision of the ontology into three separated modules is due to the fact that a correct adoption of the SAMOD methodology would have been considerably complicated by number of axioms necessary to describe a similar scenario. Consequently, SAMOD approach has been independently applied to the three different sections. However, these modules are not completely separated, rather in the final phase of the SAMOD Methodology a restricted number of classes and properties have been mutually imported across the different modules as shown in Fig. \ref{fig:zamo-mod}:

 \begin{figure}[ht!]
  \includegraphics[width=1\textwidth]{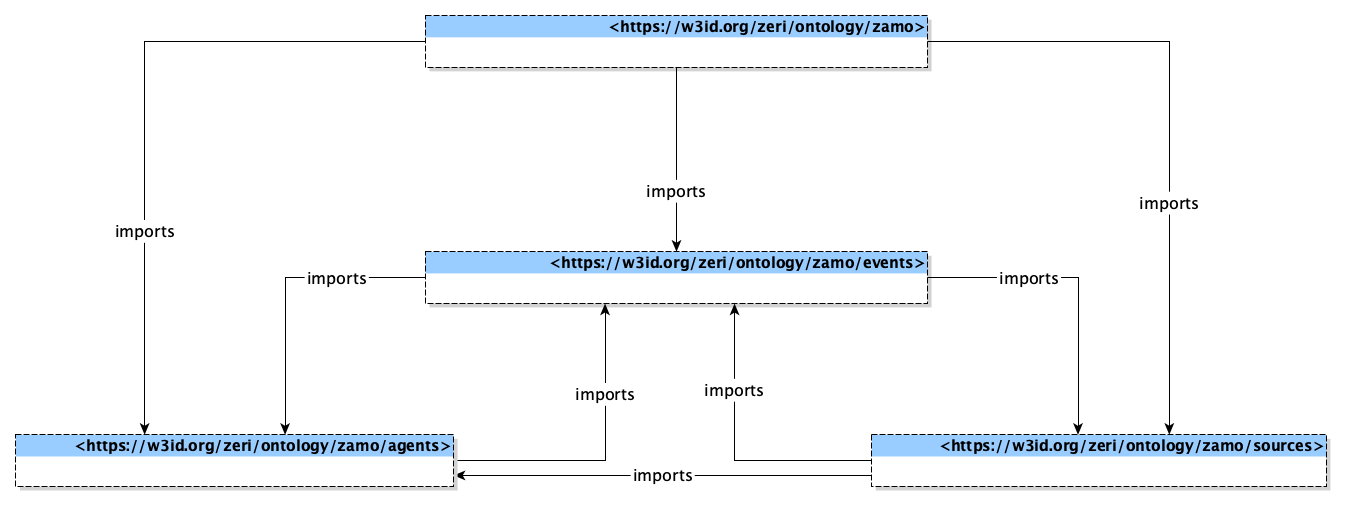}
  \caption{Conceptual model describing the import axioms of the different modules of ZAMO}
  \label{fig:zamo-mod}
\end{figure}

The preliminary analysis \parencite[see][]{veggiModellingArtMarket2023}{}{} already provides a thorough description of all SAMOD iterations for every module of the ontology. Nonetheless, during the actual development of ZAMO, some minor corrections to the guidelines described in that overview have been deemed necessary. The following subsections are hence devoted to provide a brief descriptions of the different modules in ZAMO, also mentioning their most relevant elements and modifications to the first preliminary analysis. This overview is integrated by the two final appendix: the first one lists the single scenarios and competency questions, whereas the second one contains a general overview of the alignment in a tabular format.

\subsubsection{Agents}

\begin{figure}[ht!]
  \includegraphics[width=1\textwidth]{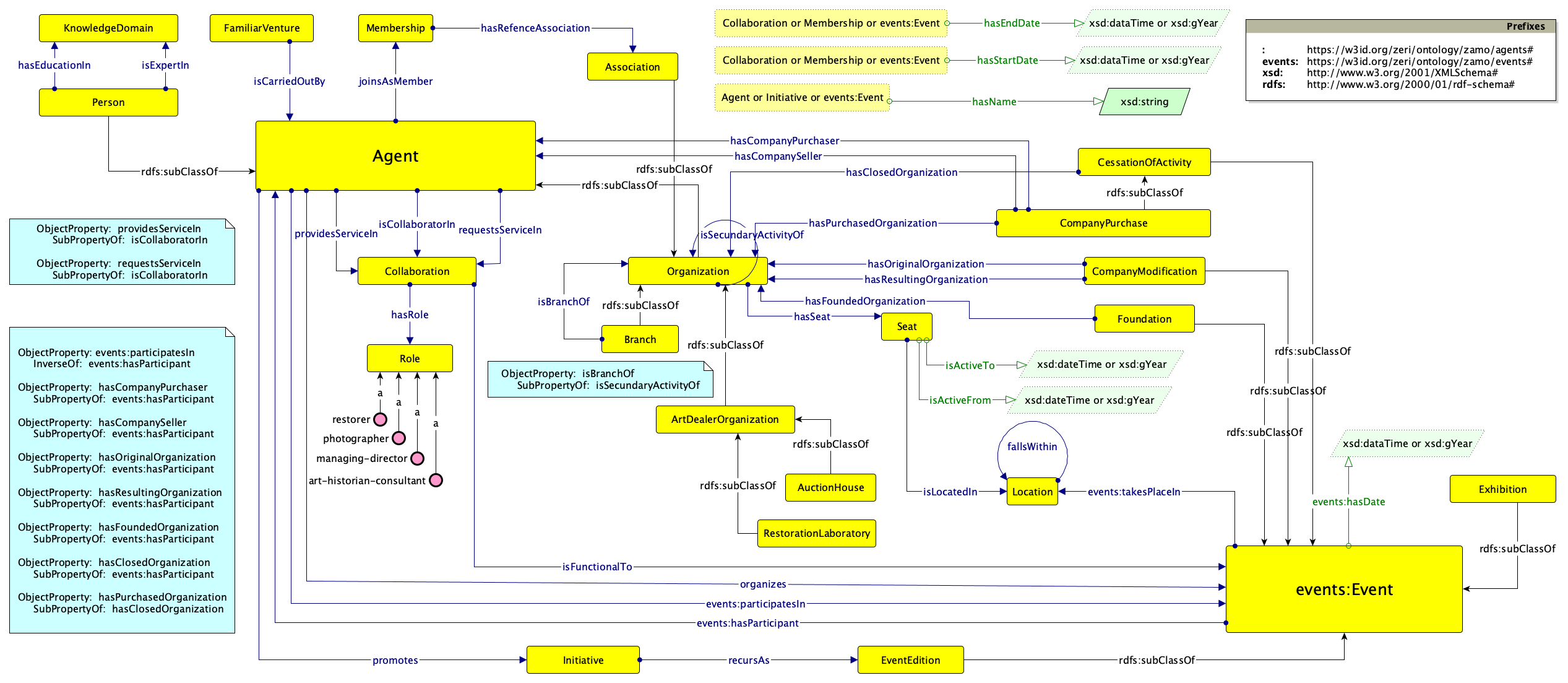}
  \caption{Conceptual model of the  the first module of ZAMO: “Agents"}
  \label{fig:zamo-agent}
\end{figure}

The first module scrutinises the most relevant actors of the art market domain, starting from the core class of “Agent", which is fully alignable with \verb|crm:E39_Actor| and \verb|foaf:Agent|, reused in the “Organization" ontology. Fig. \ref{fig:zamo-agent} contains the entirety of classes and properties asserted in this module.

This final model has been iteratively built through five iterations of the SAMOD methodology. Without any substantial change from the preliminary analysis, the first one describes how the aforementioned class of “Agent" should entail the different protagonists of the art market, i.e. both single individuals and organisations (as in CIDOC-CRM and FOAF). In addition, the first scenario indicate  as necessary a more detailed description of the class “Organization”. Even though the majority of its  subclasses  will  be  introduced  in  the  following  scenarios,  it  is  still  possible  to  introduce “ArtDealerOrganization”, which is in turn superclass of “AuctionHouse”, already defined in the ontology developed by Filipiak (see \textit{supra}). 

On the contrary, the involvement in the art market cannot be a sufficient criterion to differentiate also the entity “Person” in subclasses, since - differently from a company - an individual can be active in more than one domain and not mandatorily at the same time. “Person” class instances’ activity should be hence considered as a role and should not be defined by instantiating an individual of a specific subclass, rather it could be inferred by the participation to events (see second module). The scenario requires instead to better define individuals stating their domain of expertise ("KnowledgeDomain"). Conversely, the reference to the role within an organisation is described in the fourth scenario, which specifically deals with the different professional roles within an art company.

After this core definition of the most relevant agents in the art market, the second scenario of the module has a prime focus on the peculiarities of the class “Organization". Indeed, it defines the seat and its name, as well as specifying how different organizations may be connected by means of hierarchical relationships (as in the case of the class “Branch" and the property “isSecundaryActivityOf"). It also introduces the restauration laboratory, which is of crucial importance in the study of the art market. For example, in the case of the art dealer Sangiorgi, the presence of an \textit{ad hoc} venue to restore artefacts and create in-style reproductions was an essential sector of his antiquarian activity \parencite[see][]{mambelliPiuGrandeCentro2020}{}{}. As seen in the preliminary analysis, the activity of restoration and production does not  need  to  be  described  in this  model,  rather  it  can  be  delegated  to  CIDOC-CRM,  in particular through the classes \verb|crm:E11_Modification| and \verb|crm:E12_Production| which can be performed by any individual of type \verb|crm:E39_Actor| or its subclass.

Furthermore, comparing to the existing models, the main introduction is the concept of the familiar venture, which encompasses all the agents, both of class “Person" and “Organization", which belong to the same family across generations. The definition of this high-level entity has been deemed necessary as it has been noted that art dealer companies tend to be handed down through generations of the same family: the importance of this class permits hence scholars to observe and analyse diachronically the evolution of a familiar business, scrutinising its main protagonists.

Together with the third scenario, this iteration also provides the necessary semantics to explain the different events which may change the configuration and peculiarities of an organisation. Following the approach proposed by both CIDOC-CRM and \textit{Org}, a class has been asserted for every typology of these events: “Foundation" (\verb|E66_Formation|), “CompanyModification" (\verb|org:ChangeEvent|), and “CessationOfActivity" (\verb|crm:E68_Dissolution|), with the newly introduced subclass “CompanyPurchase"; this latter event has been interpreted as specific type of cessation of activity, as it causes a change in the reference familiar venture.

As anticipated, the forth scenario scrutinises how a person may be involved in an organisation through a professional relationship. This first version of the ontology maintained the solution individuated in the preliminary analysis, based on \textit{Org}. The key class of this scenario is indeed “Collaboration", related to \verb|org:Membership|, which refers to the fact that two agents have stipulated any form of collaboration or partnership to maximize their own interests. This can be translated also - but not mandatorily - into an employment contract. The two actors involved should be subject of a specific property stating their participation in the collaboration. This last property should be also differentiated into two sub-properties, which enable us to express who has requested and who is providing the service ("providesServiceIn" and “requestsServiceIn"). 

Following the class hierarchy of \textit{Org}, this entity should not be defined as subclass of “Event" and should relate to the definition of a role, which is expressed in turn through a controlled vocabulary. With reference to the instructions provided by the motivating scenario, a first nucleus of term shall include the managing director (also mentioned in scenario 1.1), the art historian consultant (whose domain of expertise may be rendered as in scenario 1.1 with “KnowledgeDomain"), the restorer, and the photographer. 

Lastly, each collaboration may be associated with a specific purpose: this element is harmonized with the reference CIDOC-CRM model, because the creation of catalogues and photographs, modification (i.e., restoration) of artworks can be indeed stated as individuals of types of \verb|crm:E5_Event| or its subclasses. In addition, also the activity of art historian consultant, in particular to what pertains expertise and value attribution, can be instead rendered by the subclasses of “Event", as described in the second module.

As “Collaboration" does not completely cover the polysemy of \verb|org:Membership|, the participation of a member to an association should be described in the last scenario of this module. Here, a new subclass of “Organization" is asserted, “Association" identifying connections or cooperative agreements between agents which are stipulated to defend and pursue shared professional interests. The relationship between this class and its members (of type “Agent") is mediated by a new class, “Membership”, which states a specific time interval (expressed through a data property) and has a more specific semantic comparing to \verb|org:Membership| as it excludes professional collaborations.

In addition, the scenario mentions other typology of events which characterise the activity of art dealers, such as fairs and exhibitions, which can be specified as subclasses of \verb|crm:E5_Event|. Nonetheless, a necessary distinction should be made for recurrent events: antiquarian fairs themselves can be included in this case, as each annual edition actually refers to the same series (e.g., The European Fine Art Fair, TEFAF – Maastricht). As individuated in the preliminary analysis, this distinction in two separated levels has been already tackled by DOLCE (\url{http://ontologydesignpatterns.org/wiki/Ontology:DOLCE+DnS_Ultralite}), a foundational upper-level ontology \parencite{gangemiDOLCEDnSUltraliten.d.}{}{}, with the classes \verb|dul:Collection| (for “Initiative") and \verb|dul:Event| (for “EventEdition"). Other typologies of events which directly imply the economic exchange and the definition of the monetary value are instead defined in the second module.

\subsubsection{Events}

\begin{figure}[ht!]
  \includegraphics[width=1\textwidth]{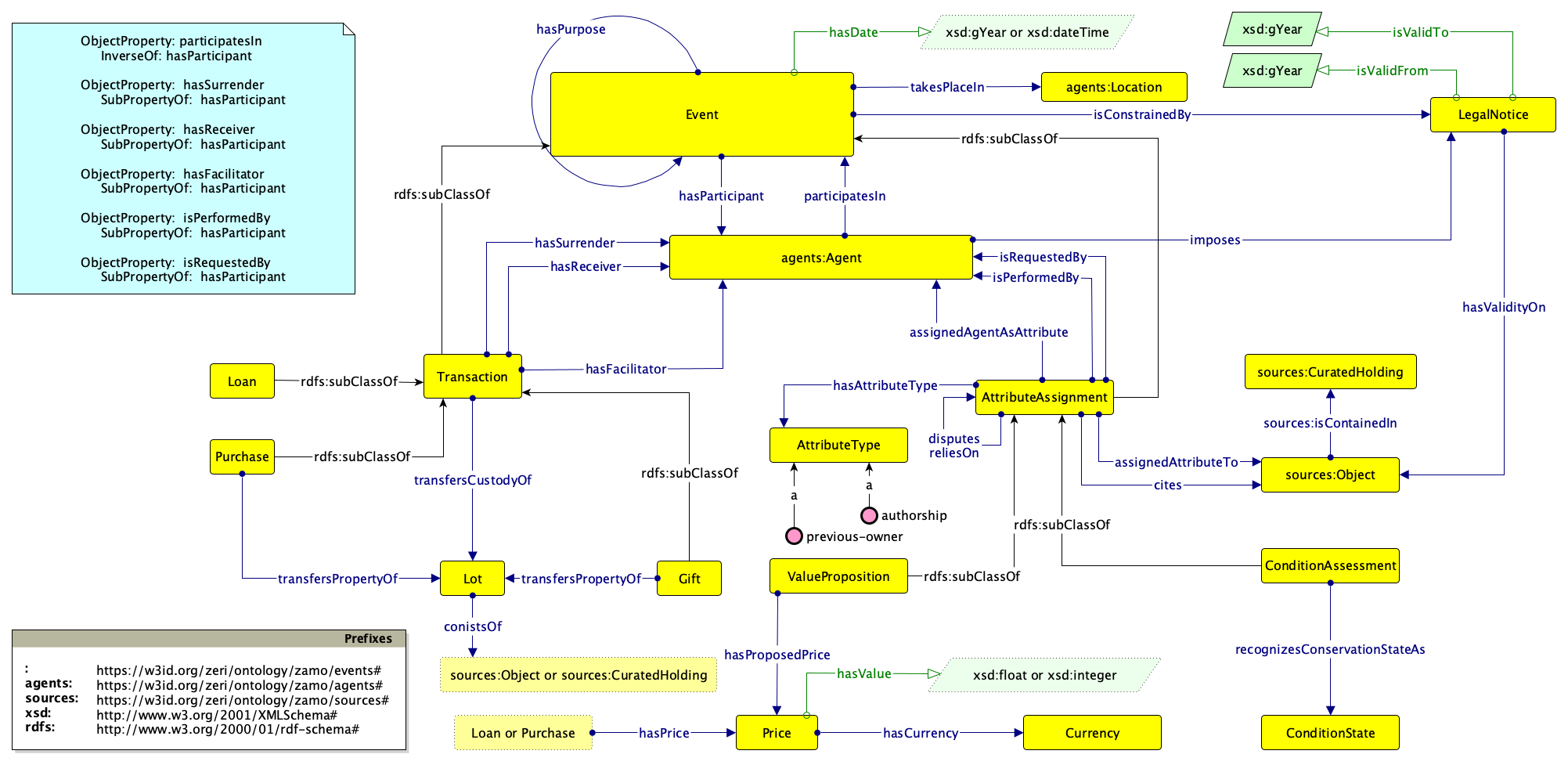}
  \caption{Conceptual model of the  the second module of ZAMO: “Events"}
  \label{fig:zamo-event}
\end{figure}

As anticipated, the analysis of the main events which characterise the art dealers' activity is dealt in the second module of ZAMO (see Fig. \ref{fig:zamo-event}), which is articulated in two main scenarios. 

The first SAMOD iteration only marginally differs from the solution proposed in the preliminary analysis. The core class is “Transaction" and has been defined as a super-class of \verb|crm:E8_Acquisition| and \verb|crm:E10_Transfer_of_Custody|, since it may results in a change of the property and/or of the custody of an artwork. It is differentiated among “Gift", “Loan" and “Purchase", which is aligned with \verb|crm:E96_Purchase|. In addition, the description of the price of the transaction has been modelled with the use of the class “Currency" and the property “hasCurrency": this streamlines the alignment with CIDOC-CRM, where \verb|crm:E98_Currency| connects a \verb|crm:E97_Monetary_Amount| to a currency of type \verb|crm:E98_Currency|.


On the contrary, for the purpose of the final alignment, some modifications were deemed necessary in the second iteration, which describes the different interpretative activities influencing the transaction. The core ones are the expertise, in which a scholar or connoisseur may make claims about an artwork authorship, previous owners or conservation state, and the value propositon.

At first, it has been proposed to model them as two separated classes. A closer look to the solutions adopted by CIDOC-CRM and HICO show instead how in the literature a different approach is encouraged. The standard ontology uses the class \verb|crm:E13_Attribute_Assignment| to render “actions of making assertions about properties of an object or any relation between two items or concepts". To better define this class, additional properties and entities have been asserted, such as \verb|crm:P35_has_identified| thanks to which it is possible to specify which element of the object is being discussed, choosing from a controlled vocabulary (a similar solution is implemented in HICO with \verb|hico:InterpretationType|). 

Whereas the expertise can be modeled as a generic instance of an “AttributeAssignment", two core subclasses have been declared. The first one, “ConditionAssessment", corresponding to CIDOC-CRM class \verb|crm:E14_Condition_Assessment|, makes possible to express claims about the conservation status of an artwork, relying on a controlled vocabulary of “ConditionState" type. The second one, instead, pertains to the “ValueProposition", for which a specific property to express the estimated monetary amount has been asserted.

\subsubsection{Sources}

\begin{figure}[ht!]
  \includegraphics[width=1\textwidth]{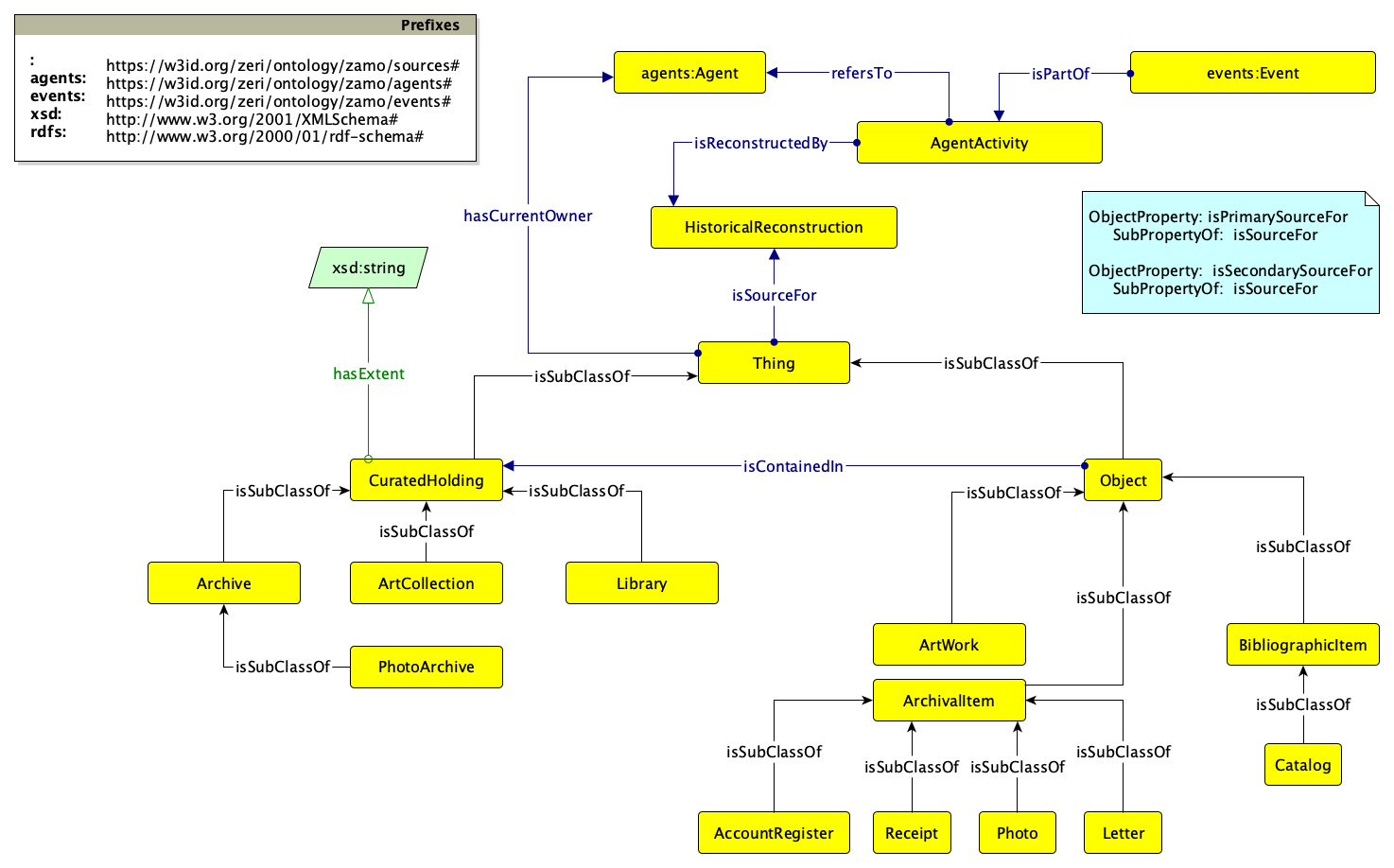}
  \caption{Conceptual model of the  the second module of ZAMO: “Events"}
  \label{fig:zamo-sources}
\end{figure}

Lastly, the third module of ZAMO, described in Fig. \ref{fig:zamo-sources}, is focused on the possible sources which may be used for the historical reconstruction of art dealers and their activity, and consists of one single SAMOD iteration.

No substantial difference has been put forward in the assertion of classes and properties. On the contrary, a new possible alignment has been serialised, which prioritises the use of the ontology ARCO. This choice is mainly justified by the fact that a full alignment with this model will be necessary in the future: ARCO is indeed a network of ontologies which is able to represent the data of the Italian Ministry of Culture. Given this purpose and being a joint initiative of National Research Council (CNR) and the ICCD, it is to become an Italian national standard for the Cultural Heritage domain \parencite[][]{carrieroArCoItalianCultural2019}{}{}.

To what pertains to the description of cultural objects, this ontology proposes new alignment possibilities comparing to the solutions outlined in the preliminary study. As a matter of fact, it contains a specific model to describe archival resource (ARCO Archive Ontology, \url{ https://w3id.org/arco/ontology/archive/}, prefix: \verb|arco-archive|): the classes “ArchivalItem" can be hence aligned with the class \verb|arco-archive:ArchivalResource|, wherease “Archive" can be expressed with the semantics of \verb|arco-archive:ArchivalCollection|. Moreover, in the ontology “ARCO Context" (\url{https://w3id.org/arco/ontology/context-description/}, prefix: \verb|arco-context|), it is possible to instantiate photo archives  - such as the one hosted by the Zeri Foundation - thanks to \verb|arco-context:PhotographicFonds|. It also proposes a data property to express the extent of the archival collection (\verb|arco-archive:extent|).

\begin{figure}[ht!]
  \includegraphics[width=1\textwidth]{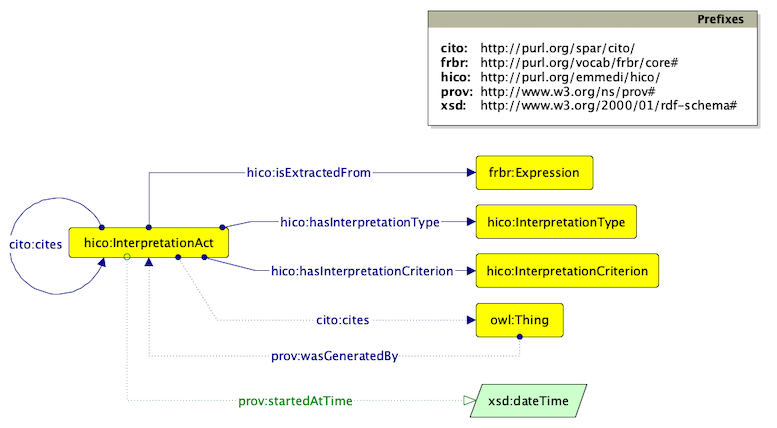}
  \caption{HICO Conceptual model, after \cite{daquinoHistoricalContextOntology2015}}
  \label{fig:hico}
\end{figure}

Instead, to harmonize the current ontology with previous experiments on the Federico Zeri Foundation \parencite[][]{daquinoEnhancingSemanticExpressivity2017}{}{}, the historical interpretation act can be aligned with the Historical Context Ontology \parencite[see][]{daquinoHistoricalContextOntology2015}{}{}, although ARCO Context proposes an alternative solution with the class \verb|arco-context:Interpretation| and the property \verb|arco-context:hasInterpretationSubject|. Indeed, in HICO, the connection of an interpretation to its subject is mediated by the use of the PROV-O ontology (\url{https://www.w3.org/TR/prov-o/}), as shown in \ref{fig:hico}. As an example (after HICO documentation), the following lines claims that a painting has been attributed to the painter Baldassare Perruzzi in Turtle serialization:

\begin{verbatim}
    @prefix crm: <http://www.cidoc-crm.org/cidoc-crm/>.
    @prefix hico: <http://purl.org/emmedi/hico/> .
    @prefix prov: <http://www.w3.org/ns/prov#> .
    
    :39794-creation-1 a crm:E65_Creation ;
        crm:P14_carried_out_by :baldassarre ;
        prov:wasGeneratedBy :39794-authorship-attribution-1.

    :39794-authorship-attribution-1 a hico:InterpretationAct. 
\end{verbatim}

\section{Discussion}

In the creation of the Art Market Ontology, major attention has been paid to the integration of existing models, such as that of Filipiak and colleagues or the Organization Ontology, which alone would not have made it possible to describe such a multifaceted domain. Furthermore, the ontology is intended to be interoperable with ICOM's standard model, CIDOC-CRM, and is also designed for future full integration with the existing RDF datasets of the Zeri Foundation, as a systematic reuse of the RDF models of the F and OA Entry has been made. In the current version, the alignment with ARCO is yet only partial.

At this first full release, some limitations can yet be identified. As a matter of fact, this project provides a first general description of the art market, able to describe the most relevant dynamics of this sector. Some minor peculiarities of a single event, of a particular agent or source might hence be omitted in the serialization of the knowledge graph, as they were not foreseen in the original motivating scenarios. For each of the three modules into which it is subdivided, one could certainly include other interactions and investigate more refined aspects and relationships linking art market-related entities.

The analysis of the research field started and remained closely linked to the contents of Federico Zeri's archive and the database that resulted from its cataloguing. In order to test its validity, it will certainly be necessary not only to take into consideration specific cases that the Foundation has deepened over the years in monographic studies, but also to apply it to datasets built by other research institutes active in the sector. 

In this perspective, opening the discussion with renown art-historical research centers mentioned in the Introduction, such as ZADIK, the Bloomsbury Digital Resources, the Getty Research Institute or The International Art Market Studies Association would be for sure be beneficial: comparing the different approaches and data structure of similar centers would allow art-historians and ontology engineers to capture with greater precision the peculiarities of this sector and could provide a number of further case studies for testing the model.

A further limitation concerns the specific nature of the sources used to explore the field of research. Catalogues, archival documents, and oral histories can sometimes deliberately conceal or distort certain details about business transactions. When collecting or using information about art market dynamics, scholars often have to deal with ambiguous, partial, or incomplete data. The ability to effectively express the possible inconsistencies, without this bias undermining the validity of future research, remains one of the goals of the model, which has yet to be tested.

\section{Conclusion}

This dissertation presents the first version of an ontology for the description the art market, ZAMO (Zeri Art Market Ontology). It stems from the increasing attention received in recent years by art market studies and from the long-standing research activities of the Federico Zeri Foundation. Beginning with the evidence collected in the Archives and Photo Library, this institution has been engaged for years in collecting information concerning mainly antiquarians active in Italy between the nineteenth and twentieth centuries. It is therefore already in possession of very refined data on some of the market protagonists that can offer a good sample to verify the correct functioning of the model. At the same time, the Foundation has recently promoted two research grants for scholars who are conducting a thorough analysis of aspects and protagonists of the market in this same chronological and temporal span. This is producing a significant amount of ex-novo data, coming from multiple sources, which must be recorded, systematized, and integrated with other resources available on the web.

Besides providing a recognition of existing conceptual models, the paper reconstructs the reference domain relying on the competency questions, examples and motivating scenarios defined by the iterations of the SAMOD Methodology. These materials have laid the basis for the identification of the most relevant entities and properties, which have been modeled both through systematic reuse of existing models and with the implementation of necessary integrations. These obseravtions have been described in a preliminary analysis \parencite[][]{veggiModellingArtMarket2023}{}{}: the dissertation hence relies on this previous study, specifying which changes have been implented in the actual development of ZAMO.

The methodological soundness of ZAMO is being assured by the full implementation of SAMOD \parencite[][]{peroniSAMODAgileMethodology2016}{}{}. In the future, it is instead essential to test this ontology with some case-tests part of the Fondazione Federico Zeri dataset, and establishing a fruitful dialogue with other research institutions that are creating authority files or other digital resources dedicated to this field.

The development of the ontology does not only satisfy the need to organize systematically the pieces of information gathered thanks to traditional methodologies of art historical research. It will hopefully show also how digital art history and knowledge organization practices can guide research even in the data collection phase, leading to the production of rich, well-structured, and integrable dataset, so to offer new strategies to further deepen the knowledge of the art market sector.

\newpage
\section*{Appendix}

\subsection*{SAMOD Scenarios and Competency Questions}\label{appendix_samod}

\par These examples enable OE to analyze the reference scenarios through practical case studies. Nonetheless, they are fictious and any reference to people, organizations and events is to be considered purely coincidental.

\subsubsection*{Module: “Agents"}


\underline{\textit{1. General overview of the main agents in the art market}}
\newline \newline
\textit{Scenario}: An art dealer activity can be carried out by either a person, specialized in a particular discipline, or an organization, which can be of various types, including an auction house, and has a person in charge (managing director). There may be people and organizations that do not carry out an art dealer activity but are nevertheless involved in the cultural industry.
\newline \newline 
\textit{Examples}:
\begin{enumerate}
    \item SR, specialized in Mannerist painting, is in charge of the auction house \textit{Antichità}.
    \item  MF, specialized in Baroque sculpture, oversees the art gallery \textit{Beaux Arts}.
    \item BV is the director of \textit{Universo mostre}, an organization which cannot be considered properly as an art dealer, albeit it operates in the artistic sector.
\end{enumerate} 

\noindent \textit{Competency Questions}:
\begin{enumerate}
    \item Who is the managing director of the auction house \textit{Antichità}? [SR]
    \item In which field is the managing director of \textit{Beaux Arts} specialized? [Baroque Sculpture]
    \item Are there organizations which are not art dealers? [BV, \textit{Universo mostre}]
\end{enumerate}

\noindent \underline{\textit{2. Possible different typologies of an art-dealer company}}
\newline \newline
\textit{Scenario}: The company is defined by a foundation, which establishes the involved actors, the name, the location, the date, and the type of activity. Several companies may refer to the same family venture, which indicates all business activities owned by the members of the same family. The company may open new branches and laboratory.
\newline \newline 
\textit{Examples}:
\begin{enumerate}
    \item The art dealer ST, after opening the \textit{Arti d'Urbe} gallery in Rome, Piazza delle Erbe, in 1949, opened a restoration laboratory in Rome, Piazza Navona, in 1956.
    \item  The art dealer CP opened a second branch of the \textit{Galleria Artimercato} in Florence in Via dei Fossi in 1977.
    \item LZ and PS founded a gallery together in Rome in 1980 under the name \textit{Arti libere}.
\end{enumerate} 

\noindent \textit{Competency Questions}:
\begin{enumerate}
    \item Which activities did ST found? [\textit{Arti d'Urbe}, 1949, Piazza delle Erbe, Rome; P.za Navona, Rome, 1956]
    \item Which are all the business activities referable to the P. family? [\textit{Artimercato}; 1977, Via dei Fossi, Firenze]
    \item Which are the activities opened in Rome between 1950 and 1985? [P.za Navona, Roma, ST, 1956; \textit{Arti Libere}, Roma, LZ e PS, 1980]
\end{enumerate}

\noindent \underline{\textit{3. Events in the life cycle of an art-dealer company}}
\newline \newline
\textit{Scenario}: The company can be changed by events (defined by a date and place), such as a change of location and names. The end of a company occurs by cessation of activity or by acquisition by another individual or company
\newline \newline 
\textit{Examples}:
\begin{enumerate}
    \item In 1910 the antiquarian RT was forced to close the gallery \textit{Tesori} in Piazza Duomo, Florence, which has been active since 1870.
    \item  The auction house owned by SD is bought by MB in 1915.
    \item LZ in 1920 changed the name of the business \textit{Arti fiorentine}, active in Florence in Via dei Fossi since 1850, to \textit{Les Arts de Paris} when it moved the company to Paris, Champs Élysées.
\end{enumerate} 

\noindent \textit{Competency Questions}:
\begin{enumerate}
    \item In which period was the RT’s organization \textit{Tesori} active? [\textit{Tesori}, 1870, 1910, Piazza Duomo, Florence]
    \item Who is the buyer of SD’s auction house? [MB, 1915]
    \item Which events modified the organization \textit{Arti fiorentine}? [LZ, 1920, \textit{Les Arts de Paris}, Paris, Champs Élysées]
\end{enumerate}

\noindent \underline{\textit{4. Professional roles and collaborations in the art market}}
\newline \newline
\textit{Scenario}: An art dealer company makes use of various professional figures, which can coincide with either  an individual or an another society, and with whom it forms collaborative ties. These relationships are defined by a time interval and are functional to an event. In addition to the person in charge, the main figures (there may be more than one) are the reference scholar (who has a background on an art-historical domain; he intervenes for attributions and catalogues), the restorer, and the photographer.
\newline \newline 
\textit{Examples}:
\begin{enumerate}
    \item The antiquarian GT has several artworks without attribution. He therefore asks AB (a leading expert in the history of the miniature) to collaborate and study these works. This collaboration is occasional and lasts from 1980 to 1982.
    \item  In order to study the works properly, collaborator AB asks for photographs. Therefore, GT asks photographer AA, with whom he collaborated from 1960 until 1985, to take the photos.
    \item FR calls three experts with different skills to draw up a catalogue: GO, a scholar of Baroque sculpture, LA, a scholar of Mannerist painting, AB, a scholar of the history of the miniature. The photographer DD also worked on the catalogue (permanent collaboration from 1950 till 1975).
    \item FZ, who had owned the company \textit{Antichità fiorentine} since 1920, died in 1948. The business was continued by his son MZ until 1970. The company was founded by FZ's uncle AQ in 1910.
\end{enumerate} 

\noindent \textit{Competency Questions}:
\begin{enumerate}
    \item With whom did the art dealer GT collaborate between 1979 and 1985? [AB, Art historian consultant, 1980, 1982; AA, Photographer, AA, 1960, 1985]
    \item  Which  photographers are active between 1960 and 2000? [AA, 1960, 1985, GT; DD, 1950, 1975, FR]
    \item Who are the art historian consultants who worked with FR to prepare his catalog? [GO, Baroque sculpture; LA, Mannerist painting; AB, History of the miniature]
    \item Who are the directors of \textit{Antichità fiorentine}? In which period were they active? [AQ, 1910, 1920; FZ, 1920, 1948; MZ, 1948, 1970]
\end{enumerate}

\noindent \underline{\textit{5. The art dealer organization as member of associations and participant to events}}
\newline \newline
\textit{Scenario}: A company may participate in several events (location, start date, end date) organized by other agents or organize one at its own premises. The main events are (but not limited to) fairs and exhibitions. Each event can be described by a catalogue. The company can be a member of an association that can promote events
\newline \newline 
\textit{Examples}:
\begin{enumerate}
    \item In 2008, the \textit{Arti belle} gallery organized an exhibition (2008-2009) in Rome. It also participates in the BIF event in Florence, organized by the association AIT in 2010.
    \item The AIT Association regularly promotes the \textit{Arti romane} fair. SA, which has been a member of AIT from 1970 to 1985, participated in this event in 1982 and 1984.
    \item The \textit{Art} gallery is a member of the AIT organization from 1977 to 1988.
\end{enumerate} 

\noindent \textit{Competency Questions}:
\begin{enumerate}
    \item To which event did Arti Belle art gallery participate? [Exhibition of \textit{Arti Belle}, Roma, 2008, 2009; BIF, AIT, Firenze, 2010]
    \item Who participated in the editions of the event \textit{Arti Romane} between 1981 and 1985? [1982, Roma, AIT, SA; 1984, Roma, AIT, SA]
    \item Who are the members of the association AIT? [SA, 1970, 1985; \textit{Art}, 1977, 1988]
\end{enumerate}

\subsubsection*{Module: “Events"}

\noindent \underline{\textit{1. Transaction typologies in the art market}}
\newline \newline
\textit{Scenario}: An agent may have ownership and custody (they do not mandatorily match) of an object or a set of artworks. An agent's main activity is the transaction, which include buying (an agent sells an asset in exchange for a sum of money; it changes possession, custody or both), lending (an agent lends an asset in exchange for a sum of money or free of charge, which changes only custody), and donating (an agent gives an asset away free of charge: in this case, both possession and custody change). The asset is defined as a lot and may be a single work or a group of works of art (sometimes even an entire collection). The transaction is identified by at least two parties: one who transfers possession and one who acquires it. A transaction may also have a facilitator.
\newline \newline 
\textit{Examples}:
\begin{enumerate}
    \item The antiquarian QC buys the work \textit{Sacra Conversazione} from the private collector ST in 1978 for a sum of €10,000..
    \item  On the occasion of the exhibition \textit{Neoclassicism} (1980), the antiquarian PP lends a terracotta by Canova to the Museo Civico in Milan.
    \item Upon TC's death (1980), his heirs donated his collection to the Museo Civico in Florence.
    \item The museum Museo Civico di Belle Arti in Rome bought the painting \textit{Birth of Bacchus} by Giulio Romano from the antiquarian RE in Florence, at the suggestion of the scholar TD
\end{enumerate} 

\noindent \textit{Competency Questions}:
\begin{enumerate}
    \item When did QC buy the painting \textit{Sacra conversazione}? [1978]
    \item Who are the clients of ST? [QC]
    \item Who is the last owner of the artworks exhibited at Neoclassicismo? [Terracotta by Canova, PP]
    \item Who participated in the acquisition of Romano’s Birth of Bacchus? [Museo Civico di Belle Arti in Rome, buyer; TD, facilitator; RE, seller]
\end{enumerate}

\noindent \underline{\textit{2. Expertise and value proposition in the art market}}
\newline \newline
\textit{Scenario}: The peculiarity of the art market is the fact that the value of an artwork cannot be deduced objectively but depends on multiple external factors. In fact, an agent can value the work independently or by relying on an expert. He reconstructs several characteristics of the work of art. Among these, some important ones are the artefact's authorship (attribution), state of preservation, collecting history (understood as a series of possessions), bibliography. Based on this analysis, an agent proposes a monetary value for the object. Both the attribution and the value proposition can be rediscussed over time by other attributive assignments: all these events influence (without necessarily coinciding) the real price of the transaction. The asset may be subject to notification (usually not withdrawn but may only be valid for a certain amount of time) by the State and this affects the value of the asset.
\newline \newline 
\textit{Examples}:
\begin{enumerate}
    \item In 2001, the antiquarian PT calls the reference scholar SB to have him study a 17th century painting (\textit{San Rocco}). He attributes the artwork to Guido Reni.
    \item  In 2005 SB studies a \textit{Deposition} attributed to Guido Reni: after some research he attributes it instead to Domenichino.
    \item The antiquarian SZ owns a painting by Tintoretto, whose value in 1980 was €15,000 because a previous appraisal showed it to be in a good state of conservation and the work is mentioned in the catalogue \textit{Le opere di Tintoretto}.
    \item Based on the analysis of Tintoretto's canvas by the scholar ML, commissioned by SZ, it was discovered that the work was owned by Count F. SZ put it on sale in 1990 at the equivalent cost of 35,000€ and at 30,000€ in 1991
    \item The work \textit{San Rocco} by Guido Reni was notified in 2008 by the Italian State
\end{enumerate} 

\noindent \textit{Competency Questions}:
\begin{enumerate}
    \item Which artworks did SB study? [\textit{San Rocco}, Guido Reni, 2001; \textit{Deposizione}, Domenichino, 2005.]
    \item Which are the different value attributions of the canvas by Tintoretto? [1980, 15000€; 1990, 35000€; 1991, 30000€]
    \item Are there artworks under notification? [\textit{San Rocco}, 2008]
\end{enumerate}

\subsubsection*{Module: “Sources"}
\noindent \underline{\textit{1. Sources for the historical reconstruction of the art market}}
\newline \newline
\textit{Scenario}: Objects may mainly (but not only) be bibliographic (comprising various types, including catalogues) or archival item (such as, but not limited to, account ledgers, receipts, letters, photos) and works of art; they may be contained within a curated holding (library, archive, including photo archive, or art collection; they may have extension indications) which is identified by a current owner (an agent) and which may in turn be contained by another holding.These elements can be used as sources for events or agents: some of them are precipitates of the agent's activity (primary source), while others are the result of later studies (secondary source).
\newline \newline 
\textit{Examples}:
\begin{enumerate}
    \item The archival collection of the antiquarian GT, comprising 600 photos and 300 letters from 1940 to 1990, is now owned by the SZ Foundation.
    \item  The \textit{Art Institute} research center owns KG's archive, which contains several primary sources, such as receipt \#KG1950 from the sale of a predella by Gentile da Fabriano and a letter to SZ.
    \item There is also KG's autobiography \textit{My life}, the unpublished manuscript of which is now in the possession of the heir FG.
    \item There are also secondary sources on KG: the volume \textit{KG: An Art Collectionist} and the website \textit{KG on the web}.
\end{enumerate} 

\noindent \textit{Competency Questions}:
\begin{enumerate}
    \item Who is the owner of GT’s archive? [Foundation SZ]
    \item Which are the sources which can be used to reconstruct the life and the activity of KG? [KG’s archive, Art Institute; Receipt \#KG1950, Art Institute; Letter to SZ, Art Institute; \textit{KG. My life}, FG; \textit{KG: An Art Collectionist}; \textit{KG on the web}]
\end{enumerate}
\newpage

\newgeometry{left=1.5cm, right=1.5cm}

\subsection*{Alignment Tables}
\label{appendix_alignment}

This section provides an alignment of ZAMO classes and properties in a tabular format. These tables use the following prefixes:

\begin{itemize}
    \item SKOS (\verb|skos|): \verb|http://www.w3.org/2004/02/skos/core#|
    \item CIDOC-CRM (\verb|crm|): \verb|http://www.cidoc-crm.org/cidoc-crm/|
    \item ARCO, Context Description (\verb|arco-context|): \verb|https://w3id.org/arco/ontology/context-description/|
    \item ARCO, Archive Ontology (\verb|arco-archive|): \verb|https://w3id.org/arco/ontology/archive/|
    \item Organization Ontology (\verb|org|): \verb|http://www.w3.org/ns/org#|
    \item DOLCE (\verb|dul|): \verb|http://www.loa-cnr.it/ontologies/DUL.owl#|
    \item F-Entry (\verb|fentry|): \verb|http://www.essepuntato.it/2014/03/fentry/|
    \item HICO (\verb|hico|): \verb|http://purl.org/emmedi/hico/| 
    \item PROV-O (\verb|prov|): \verb|http://www.w3.org/ns/prov#|
    \item FABIO (\verb|fabio|): \verb|http://purl.org/spar/fabio/|
    \item CITO (\verb|cito|): \verb|http://purl.org/spar/cito/|
\end{itemize}

\subsubsection*{Module: “Agents"} 

\noindent Base URI (for column “ZAMO URI"): \verb|https://w3id.org/zeri/ontology/zamo/agents#|

\begin{longtable}{| p{0.35\textwidth} | p{0.2\textwidth} | p{0.37\textwidth} |} 
    \hline
        \rowcolor[HTML]{C0C0C0} 
        \textbf{Zamo URI} & \textbf{Skos Property} & \textbf{Aligned URI}\\ \hline
        
        \verb|Agent| & \verb|skos:exactMatch| & \verb|crm:E39_Actor| \\ \hline
        \verb|Location| & \verb|skos:exactMatch| & \verb|crm:E53_Place| \\ \hline
        \verb|Seat| & \verb|skos:exactMatch| & \verb|org:Site| \\ \hline
        \verb|Organization| & \verb|skos:exactMatch| & \verb|org:Organization| \\ \hline
        \verb|CompanyModification| & \verb|skos:exactMatch| & \verb|org:ChangeEvent| \\ \hline
        \verb|Role| & \verb|skos:relatedMatch| & \verb|org:Role| \\ \hline
        \verb|Collaboration| & \verb|skos:relatedMatch| & \verb|org:Membership| \\ \hline
        \verb|KnowledgeDomain| & \verb|skos:narrowMatch| & \verb|fabio:SubjectDiscipline| \\ \hline
        \verb|Initiative| & \verb|skos:narrowMatch| & \verb|dul:Collection| \\ \hline
        \verb|EventEdition| & \verb|skos:narrowMatch| & \verb|dul:Event| \\ \hline
        \verb|Foundation| & \verb|skos:exactMatch| & \verb|crm:E66_Formation| \\ \hline
        \verb|Branch| & \verb|skos:narrowMatch| & \verb|org:OrganizationalUnit| \\ \hline
        \verb|CessationOfActivity| & \verb|skos:exactMatch| & \verb|crm:E68_Dissolution| \\ \hline
        \verb|Person| & \verb|skos:exactMatch| & \verb|crm:E21_Person| \\ \hline
        \verb|hasResultingOrganization| & \verb|skos:exactMatch| & \verb|org:resultingOrganization| \\ \hline
        \verb|hasOriginalOrganization| & \verb|skos:exactMatch| & \verb|org:originalOrganization| \\ \hline
        \verb|hasFoundedOrganization| & \verb|skos:exactMatch| & \verb|crm:P95_has_formed| \\ \hline
        \verb|isBranchOf| & \verb|skos:narrowMatch| & \verb|org:hasUnit| \\ \hline
        \verb|hasRole| & \verb|skos:relatedMatch| & \verb|org:role| \\ \hline
        \verb|fallsWithin| & \verb|skos:exactMatch| & \verb|crm:P89_falls_within| \\ \hline
        \verb|isLocatedIn| & \verb|skos:closeMatch| & \verb|org:siteAddress| \\ \hline
        \verb|isSecundaryActivityOf| & \verb|skos:narrowMatch| & \verb|org:subOrganizationOf| \\ \hline
        \verb|hasSeat| & \verb|skos:exactMatch| & \verb|org:hasSite| \\ \hline
        \verb|hasClosedOrganization| & \verb|skos:exactMatch| & \verb|crm:P99_dissolved| \\ \hline
        \verb|isCollaboratorIn| & \verb|skos:relatedMatch| & \verb|org:hasMember| \\ \hline
\end{longtable}

\subsubsection*{Module: “Events"}

\noindent Base URI (for column “ZAMO URI"): \verb|https://w3id.org/zeri/ontology/zamo/events#|

\begin{longtable}{| p{0.35\textwidth} | p{0.2\textwidth} | p{0.37\textwidth} |} 
    \hline
        \rowcolor[HTML]{C0C0C0} 
        \textbf{Zamo URI} & \textbf{Skos Property} & \textbf{Aligned URI}\\ \hline
        
        \verb|Price| & \verb|skos:exactMatch| & \verb|crm:E97_Monetary_Amount| \\ \hline
        \verb|Transaction| & \verb|skos:broaderMatch| & \verb|crm:E10_Transfer_of_Custody| \\ \hline
        \verb|Transaction| & \verb|skos:broaderMatch| & \verb|crm:E8_Acquisition| \\ \hline
        \verb|Purchase| & \verb|skos:exactMatch| & \verb|crm:E96_Purchase| \\ \hline
        \verb|AttributeAssignment| & \verb|skos:narrowMatch| & \verb|crm:E13_Attribute_Assignment| \\ \hline
        \verb|Event| & \verb|skos:exactMatch| & \verb|crm:E5_Event| \\ \hline
        \verb|ConditionAssessment| & \verb|skos:exactMatch| & \verb|crm:E3_Condition_State| \\ \hline
        \verb|LegalNotice| & \verb|skos:narrowMatch| & \verb|crm:E89_Propositional_Object| \\ \hline
        \verb|AttributeType| & \verb|skos:narrowMatch| & \verb|crm:E55_Type| \\ \hline
        \verb|ConditionState| & \verb|skos:exactMatch| & \verb|crm:E3_Condition_State| \\ \hline
        \verb|Currency| & \verb|skos:exactMatch| & \verb|crm:E98_Currency| \\ \hline
        \verb|ValueProposition| & \verb|skos:narrowMatch| & \verb|crm:E16_Measurement| \\ \hline
        \verb|hasCurrency| & \verb|skos:exactMatch| & \verb|crm:P180_has_currency| \\ \hline
        \verb|assignsAttributeTo| & \verb|skos:narrowMatch| & \verb|crm:P140_assigned_attribute_to| \\ \hline
        \verb|transfersCustodyOf| & \verb|skos:exactMatch| & \verb|crm:P30_transferred_custody_of| \\ \hline
        \verb|disputes| & \verb|skos:narrowMatch| & \verb|cito:disputes| \\ \hline
        \verb|hasSurrender| & \verb|skos:broaderMatch| & \verb|crm:P23_transferred_title_from| \\ \hline
        \verb|hasSurrender| & \verb|skos:broaderMatch| & \verb|crm:P28_custody_surrendered_by| \\ \hline
        \verb|takesPlaceIn| & \verb|skos:exactMatch| & \verb|crm:P7_took_place_at| \\ \hline
        \verb|participatesIn| & \verb|skos:exactMatch| & \verb|crm:P11i_participated_in| \\ \hline
        \verb|isPerformedBy| & \verb|skos:narrowMatch| & \verb|crm:P14_carried_out_by| \\ \hline
        \verb|hasReceiver| & \verb|skos:broaderMatch| & \verb|crm:P22_transferred_title_to| \\ \hline
        \verb|hasReceiver| & \verb|skos:broaderMatch| & \verb|crm:P29_custody_received_by| \\ \hline
        \verb|hasPrice| & \verb|skos:exactMatch| & \verb|crm:P179_had_sales_price| \\ \hline
        \verb|isConstrainedBy| & \verb|skos:narrowMatch| & \verb|crm:P15_was_influenced_by| \\ \hline
        \verb|hasProposedPrice| & \verb|skos:narrowMatch| & \verb|crm:P40_observed_dimension| \\ \hline
        \verb|transfersPropertyOf| & \verb|skos:broaderMatch| & \verb|crm:P24_transferred_title_of| \\ \hline
        \verb|assignsAgentAsAttribute| & \verb|skos:narrowMatch| & \verb|crm:P141_assigned| \\ \hline
        \verb|consistsOf| & \verb|skos:narrowMatch| & \verb|dul:hasPart| \\ \hline
        \verb|cites| & \verb|skos:narrowMatch| & \verb|cito:cites| \\ \hline
        \verb|hasPurpose| & \verb|skos:exactMatch| & \verb|crm:P20_had_specific_purpose| \\ \hline
        \verb|reliesOn| & \verb|skos:narrowMatch| & \verb|cito:isSupportedBy| \\ \hline
        \verb|recognizesConservationStatusAs| & \verb|skos:exactMatch| & \verb|crm:P35_has_identified| \\ \hline
        \verb|hasParticipant| & \verb|skos:exactMatch| & \verb|crm:P11_had_participant| \\ \hline
        \verb|hasAttributeType| & \verb|skos:exactMatch| & \verb|crm:P177_assigned_property_of_type| \\ \hline
\end{longtable}

\subsubsection*{Module: “Sources"}

\noindent Base URI (for column “ZAMO URI"): \verb|https://w3id.org/zeri/ontology/zamo/sources#|

\begin{longtable}{| p{0.35\textwidth} | p{0.2\textwidth} | p{0.37\textwidth} |} 
    \hline
        \rowcolor[HTML]{C0C0C0} 
        \textbf{Zamo URI} & \textbf{Skos Property} & \textbf{Aligned URI}\\ \hline
        
        \verb|HistoricalReconstruction| & \verb|skos:narrowMatch| & \verb|hico:InterpretationAct| \\ \hline
        \verb|Thing| & \verb|skos:exactMatch| & \verb|crm:E18_Physical_Thing| \\ \hline
        \verb|CuratedHolding| & \verb|skos:exactMatch| & \verb|crm:E78_Curated_Holding| \\ \hline
        \verb|ArchivalItem| & \verb|skos:exactMatch| & \verb|arco-archive:ArchivalResource| \\ \hline
        \verb|Letter| & \verb|skos:exactMatch| & \verb|fabio:Letter| \\ \hline
        \verb|Object| & \verb|skos:exactMatch| & \verb|crm:E22_Human-Made_Object| \\ \hline
        \verb|Artwork| & \verb|skos:exactMatch| & \verb|fabio:ArtisticWork| \\ \hline
        \verb|Photo| & \verb|skos:exactMatch| & \verb|fentry:Photograph| \\ \hline
        \verb|Archive| & \verb|skos:exactMatch| & \verb|arco-archive:ArchivalCollection| \\ \hline
        \verb|PhotoArchive| & \verb|skos:narrowMatch| & \verb|arco-context:PhotographicFonds| \\ \hline
        \verb|BibliographicItem| & \verb|skos:narrowMatch| & \verb|fabio:Item| \\ \hline
        \verb|Catalog| & \verb|skos:exactMatch| & \verb|fabio:Catalog| \\ \hline
        \verb|isContainedIn| & \verb|skos:narrowMatch| & \verb|crm:P46i_forms_part_of| \\ \hline
        \verb|isSourceFor| & \verb|skos:narrowMatch| & \verb|cito:isDocumentedBy| \\ \hline
        \verb|hasCurrentOwner| & \verb|skos:exactMatch| & \verb|crm:P52_has_current_owner| \\ \hline
        \verb|isReconstructedBy| & \verb|skos:narrowMatch| & \verb|prov:wasGeneratedBy| \\ \hline
        \verb|hasExtent| & \verb|skos:exactMatch| & \verb|arco-archive:extent| \\ \hline
\end{longtable}

\restoregeometry

\newpage
\printbibliography

@inproceedings{doerrDocumentingEventsMetadata2006,
	title = {Documenting Events in Metadata},
	booktitle = {{VAST}: International Symposium on Virtual Reality, Archaeology and Intelligent Cultural Heritage},
	publisher = {The Eurographics Association},
	author = {Doerr, Martin and Kritsotaki, Heraklion},
	editor = {Ioannides, Marinos and Arnold, David and Niccolucci, Franco and Mania, Katerina},
	date = {2006},
}

@incollection{filipiakQuantitativeAnalysisArt2016,
	location = {Cham},
	title = {Quantitative Analysis of Art Market Using Ontologies, Named Entity Recognition and Machine Learning: A Case Study},
	volume = {255},
	pages = {79--90},
	booktitle = {Business Information Systems},
	publisher = {Springer International Publishing},
	author = {Filipiak, Dominik and Agt-Rickauer, Henning and Hentschel, Christian and Filipowska, Agata and Sack, Harald},
	editor = {Abramowicz, Witold and Alt, Rainer and Franczyk, Bogdan},
	date = {2016},
	doi = {10.1007/978-3-319-39426-8_7},
	note = {Series Title: Lecture Notes in Business Information Processing},
}

@inproceedings{guizzardiCoreOntologyEconomic2020,
	title = {A Core Ontology for Economic Exchanges},
	pages = {364--374},
	booktitle = {Conceptual Modeling: 39th International Conference, {ER} 2020, Vienna, Austria, November 3–6, 2020, Proceedings.},
	author = {Guizzardi, Giancarlo and Amaral, Glenda and Porello, Daniele and Prince Sales, Tiago},
	date = {2020-10},
}

@online{reynoldsOrganizationOntology2014,
	location = {W3C Recommendation},
	title = {The Organization Ontology},
	url = {https://www.w3.org/TR/vocab-org/},
	author = {Reynolds, Dave},
	urldate = {2023-03-16},
	date = {2014},
}

@article{daquinoEnhancingSemanticExpressivity2017,
	title = {Enhancing Semantic Expressivity in the Cultural Heritage Domain: Exposing the Zeri Photo Archive as Linked Open Data},
	volume = {10},
	issn = {1556-4673, 1556-4711},
	doi = {10.1145/3051487},
	shorttitle = {Enhancing Semantic Expressivity in the Cultural Heritage Domain},
	pages = {1--21},
	number = {4},
	journaltitle = {Journal on Computing and Cultural Heritage},
	shortjournal = {J. Comput. Cult. Herit.},
	author = {Daquino, Marilena and Mambelli, Francesca and Peroni, Silvio and Tomasi, Francesca and Vitali, Fabio},
	date = {2017-10-26},
	langid = {english},
}

@article{peroniFaBiOCiTOOntologies2012,
	title = {{FaBiO} and {CiTO}: Ontologies for describing bibliographic resources and citations},
	volume = {17},
	issn = {15708268},
	doi = {10.1016/j.websem.2012.08.001},
	shorttitle = {{FaBiO} and {CiTO}},
	pages = {33--43},
	journaltitle = {Journal of Web Semantics},
	shortjournal = {Journal of Web Semantics},
	author = {Peroni, Silvio and Shotton, David},
	urldate = {2023-03-16},
	date = {2012-12},
	langid = {english},
}

@incollection{daquinoHistoricalContextOntology2015,
	location = {Cham},
	title = {Historical Context Ontology ({HiCO}): A Conceptual Model for Describing Context Information of Cultural Heritage Objects},
	volume = {544},
	shorttitle = {Historical Context Ontology ({HiCO})},	pages = {424--436},
	booktitle = {Metadata and Semantics Research},
	publisher = {Springer International Publishing},
	author = {Daquino, Marilena and Tomasi, Francesca},
	editor = {Garoufallou, Emmanouel and Hartley, Richard J. and Gaitanou, Panorea},
	date = {2015},
	doi = {10.1007/978-3-319-24129-6_37},
	note = {Series Title: Communications in Computer and Information Science},
}

@inproceedings{peroniSAMODAgileMethodology2016,
	title = {{SAMOD}: an agile methodology for the development of ontologies},
	pages = {1--14},
	booktitle = {Proceedings of the 13th {OWL}: Experiences and Directions Workshop and 5th {OWL} reasoner evaluation workshop ({OWLED}-{ORE} 2016)},
	author = {Peroni, Silvio},
	date = {2016},
}

@article{poveda2014oops,
	title = {{OOPS}! ({OntOlogy} pitfall scanner!): An on-line tool for ontology evaluation},
	volume = {10},
	pages = {7--34},
	number = {2},
	journaltitle = {International Journal on Semantic Web and Information Systems ({IJSWIS})},
	author = {Poveda-Villalón, María and Gómez-Pérez, Asunción and Suárez-Figueroa, Mari Carmen},
	date = {2014},
	note = {Publisher: {IGI} Global},
}

@inproceedings{gangemiDiTTODiagramsTransformation2013,
	title = {{DiTTO}: Diagrams transformation into {OWL}},
	volume = {1035},
	url = {https://www.scopus.com/inward/record.uri?eid=2-s2.0-84924705694&partnerID=40&md5=479affbb65f3508aab864416b1dc2dca},
	series = {{CEUR} Workshop Proceedings},
	eventtitle = {12th International Semantic Web Conference, {ISWC} 2013},
	pages = {5 -- 8},
	booktitle = {Proceedings of the {ISWC} 2013 Posters \& Demonstrations Track},
	author = {Gangemi, Aldo and Peroni, Silvio},
	date = {2013},
}

@incollection{falcoModellingOWLOntologies2014,
	location = {Cham},
	title = {Modelling {OWL} Ontologies with Graffoo},
	volume = {8798},
	pages = {320--325},
	booktitle = {The Semantic Web: {ESWC} 2014 Satellite Events},
	publisher = {Springer International Publishing},
	author = {Falco, Riccardo and Gangemi, Aldo and Peroni, Silvio and Shotton, David and Vitali, Fabio},
	editor = {Presutti, Valentina and Blomqvist, Eva and Troncy, Raphael and Sack, Harald and Papadakis, Ioannis and Tordai, Anna},
	date = {2014},
	langid = {english},
	doi = {10.1007/978-3-319-11955-7_42},
	note = {Series Title: Lecture Notes in Computer Science},
}

@inbook{mambelliPiuGrandeCentro2020,
	location = {Cinisello Balsamo},
	title = {"Il più grande centro commerciale di oggetti d’arte": la galleria Sangiorgi tra strategie di marketing e artigianato artistico,},
	booktitle = {Capitale e crocevia. Il mercato dell’arte nella Roma sabauda},
	publisher = {Silvana Editoriale},
	author = {Mambelli, Francesca},
	bookauthor = {Bacchi, Andrea and Capitelli, Giovanna},
	date = {2020},
}

@online{gangemiDOLCEDnSUltraliten.d.,
	title = {{DOLCE}+{DnS} Ultralite Documentation, v. 3.34},
	url = {http://www.ontologydesignpatterns.org/ont/dul/DUL.owl},
	author = {Gangemi, Aldo},
	urldate = {2023-05-21},
}

@inproceedings{mambelliRisorsaOnlineStoria2014,
	location = {Roma},
	title = {Una risorsa online per la storia dell'arte: il database della Fondazione Federico Zeri},
	eventtitle = {Primo convegno annuale dell'{AIUCD}},
	pages = {113--125},
	booktitle = {Digital Humanities. Progetti italiani ed esperienze di convergenza multidisciplinare},
	publisher = {Sapienza Università Editrice},
	author = {Mambelli, Francesca},
	date = {2014},
}

@article{caraffaIntroductionPhotographyArt2020,
	title = {Introduction. Photography, Art, Market, and the production of value},
	volume = {62},
	pages = {3--9},
	number = {1},
	journaltitle = {Mitteilungen des Kunsthistorischen Institutes in Florenz},
	author = {Caraffa, Costanza and Bärnighausen, Julia},
	date = {2020},
}

@online{bloomsburyBloomsburyArtMarketsn.d.,
	title = {Bloomsbury Art Markets. Protagonists, Networks, Provenances},
	url = {https://www.artmarketdictionary.com/},
	author = {{Bloomsbury}},
	urldate = {2023-08-30},
}

@book{bacchi2020capitale,
	title = {Capitale e crocevia. Il mercato dell'arte nella Roma sabauda},
	volume = {8},
	publisher = {Fondazione Federico Zeri-Silvana Editoriale},
	author = {Bacchi, Andrea and Capitelli, Giovanna},
	date = {2020},
}

@article{veggiModellingArtMarket2023,
	title = {Modelling {The} {Art} {Market} in {The} {Semantic} {Web}. {A} {Preliminary} {Analysis}},
	copyright = {Creative Commons Attribution 4.0 International},
	doi = {10.6092/ISSN.2532-8816/17208},
	journal = {Umanistica Digitale},
	author = {Veggi, Manuele and Mambelli, Francesca},
	month = dec,
	year = {2023},
	pages = {141--166},
}

@misc{owlworkinggroupPunning2007,
	title = {Punning},
	url = {https://www.w3.org/2007/OWL/wiki/PropertyPunning.html},
	author = {{OWL Working Group}},
	year = {2007},
}

@article{grauOWLNextStep2008,
	title = {{OWL} 2: {The} next step for {OWL}},
	volume = {6},
	issn = {15708268},
	shorttitle = {{OWL} 2},
	url = {https://linkinghub.elsevier.com/retrieve/pii/S1570826808000413},
	doi = {10.1016/j.websem.2008.05.001},
	language = {en},
	number = {4},
	urldate = {2024-02-17},
	journal = {Journal of Web Semantics},
	author = {Grau, Bernardo Cuenca and Horrocks, Ian and Motik, Boris and Parsia, Bijan and Patel-Schneider, Peter and Sattler, Ulrike},
	month = nov,
	year = {2008},
	pages = {309--322},
}

@inproceedings{lamparter2007preference,
  title={Preference-based selection of highly configurable web services},
  author={Lamparter, Steffen and Ankolekar, Anupriya and Studer, Rudi and Grimm, Stephan},
  booktitle={Proceedings of the 16th international conference on World Wide Web},
  pages={1013--1022},
  year={2007}
}

@incollection{carrieroArCoItalianCultural2019,
	address = {Cham},
	title = {{ArCo}: {The} {Italian} {Cultural} {Heritage} {Knowledge} {Graph}},
	volume = {11779},
	shorttitle = {{ArCo}},
	url = {https://link.springer.com/10.1007/978-3-030-30796-7_3},
	language = {en},
	urldate = {2024-02-18},
	booktitle = {The {Semantic} {Web} – {ISWC} 2019},
	publisher = {Springer International Publishing},
	author = {Carriero, Valentina Anita and Gangemi, Aldo and Mancinelli, Maria Letizia and Marinucci, Ludovica and Nuzzolese, Andrea Giovanni and Presutti, Valentina and Veninata, Chiara},
	editor = {Ghidini, Chiara and Hartig, Olaf and Maleshkova, Maria and Svátek, Vojtěch and Cruz, Isabel and Hogan, Aidan and Song, Jie and Lefrançois, Maxime and Gandon, Fabien},
	year = {2019},
	doi = {10.1007/978-3-030-30796-7_3},
	note = {Series Title: Lecture Notes in Computer Science},
	pages = {36--52},
}

@inproceedings{peroni2012making,
  title={Making Ontology Documentation with LODE.},
  author={Peroni, Silvio and Shotton, David M and Vitali, Fabio and others},
  booktitle={I-SEMANTICS (Posters \& Demos)},
  pages={63--67},
  year={2012}
}
\end{document}